\documentstyle[psfig]{elsart}

       \newcommand{\GeV}{\,\mathrm{GeV}}

\newcommand{\gray}{$\gamma$-ray\ } \newcommand{\grays}{ $\gamma$-rays\
}

\begin{document} 

\begin{frontmatter}

\title{The Likely Cause of the EGRET GeV Anomaly and its Implications}

\author{F.W. Stecker}

\address{NASA  Goddard  Space  Flight  Center,  Greenbelt,  MD  20771}
\author{S.     D.    Hunter}    
\address{NASA Goddard  Space Flight Center, Greenbelt,  MD 20771, USA}
\author{D.      A.     Kniffen}    
\address{NASA Goddard  Space Flight Center, Greenbelt,  MD 20771, USA}
\address{University  Space Research  Association, Columbia,  MD 21044,
USA}

\begin{abstract}

Analysis of  data from the EGRET  \gray detector on  the Compton Gamma
Ray Observatory indicated an anomaly  in the form of an excess diffuse
galactic  flux  at GeV  energies  over  that  which was  theoretically
predicted. Various explanations for  this anomaly have been put forth,
including the  invocation of supersymmetric  dark matter annihilation.
We reexamine  these explanations here,  including a new  discussion of
the possible systematic errors in the sensitivity determination of the
EGRET detector.  We  conclude that the most likely  explanation of the
EGRET  ``GeV anomaly'' was  an error  in the  estimation of  the EGRET
sensitivity at energies above $\sim$1  GeV. We give reasons why such a
situation could have  occured.  We find evidence from  our new all-sky
analysis which  is inconsistent with  the assumption that  the anomaly
can be a  signal of supersymmetric dark matter  annihilation.  We also
reconfirm the original results of  the EGRET team on the extragalactic
\gray background  spectrum.  There  are important implications  of our
analysis  for the  upcoming  Gamma Ray  Large  Area Telescope  (GLAST)
mission.
\end{abstract}

\end{frontmatter}

\section{Introduction}

The EGRET \gray  detector, a spark chamber telescope  flown aboard the
Compton   Gamma   Ray   Observatory   satellite,   provided   detailed
measurements   of  astrophysical  \gray   sources  and   galactic  and
extragalactic diffuse \gray fluxes. The reported galactic diffuse flux
was measured and mapped over the whole sky.

Theoretical studies of the  physics and astrophysics of galactic \gray
production  provided predictions  of  the expected  fluxes and  energy
spectra  \cite{st77}.  In  the energy  range above  $\sim$1 GeV  , the
fluxes reported  by the EGRET team  were up to  $\sim$60\% higher than
the  theoretical predictions  \cite{hu97}.  This  apparent discrepancy
between the  reported fluxes and the theoretical  calculations will be
referred to here as the ``GeV anomaly''.

Unresolved galactic point  sources can be ruled out  as an explanation
for  the GeV  anomaly  for two  reasons:  (1) they  would be  strongly
concentrated  in the galactic  plane and  the GE  anomaly, as  we will
show, is  seen isotropically over the  whole sky, and  (2) the largest
class  of galactic  point sources  are  pulsars and  such sources  are
concentrated in  the inner galaxy and  they make up less  than 15\% of
the total galactic flux \cite{ha81}.

There have been three approaches for accounting for the GeV anomaly as
a  diffuse phenomenon, {\it  viz.}  either  (1) invoking  a cosmic-ray
electron  source spectrum  proportional to  $E^{-2}$ with  a resulting
significant Compton  \gray ~component above 1 GeV  increasing the flux
\cite{po98},  (2) making  modifications to  the  assumptions regarding
both the primary cosmic-ray  nucleon and electron spectra in numerical
models  in  order to  push  up  the  total theoretical  \gray  ~fluxes
\cite{st04a},  or (3) postulating  new physics,  namely supersymmetric
dark matter annihilation, to account for the anomaly \cite{de05}.

The first  approach postulates  that the cosmic-ray  electron spectrum
observed at  Earth is  much steeper than  the average spectrum  in the
Galaxy, but that  this {\it ad hoc} situation can  result as an effect
of the distribution of the  supernova remnants (SNR) which produce the
electrons and electron propagation effects. A prediction of this model
is a  center-anticenter asymmetry in  the anomaly owing to  the strong
galactocentric  radial distribution of  SNR.  A  reduced bump  at high
galactic latitudes would also be  expected, owing to the steeper local
cosmic-ray electron spectrum.

Approaches (2)  and (3) have significant  implications. Increasing the
galactic  \gray  production rate  at  GeV  energies  in the  model  of
Ref.~\cite{st04a}  produced a reduction  in the  implied extragalactic
diffuse flux  in the followup calculation~\cite{st04b},  with a marked
dip at energies  near $\sim 1 \GeV$.  This is  opposed to the original
determination of  the extragalactic  background spectrum by  the EGRET
team \cite{sr98}.   Postulating that  the GeV anomaly  is caused  by a
\gray component  from supersymmetric dark matter  annihilation has, of
course,  much  greater implications.   It  is  therefore important  to
reexamine the issue of the  GeV anomaly by paying careful attention to
the   collateral  implications  of   deviations  from   the  canonical
predictions \cite{st77}  and to  examine the more  prosaic possibility
that the assumed high-energy sensitivity of the EGRET detector may not
have been correct. In this paper, we will make the case that a problem
in  the analysis  of the  EGRET  sensitivity calibration  is the  most
likely explanation  for the ``GeV  anomaly'' and that the  dark matter
hypothesis can  be ruled  out by an  examination of the  ~GeV spectrum
over the whole sky.

\section{Production of Diffuse Galactic \grays}

Galactic \grays are produced by interactions of relativistic electrons
and protons with interstellar  gas and photons. The physical processes
involved  are electron  bremsstrahlung, neutral  pion  production, and
Compton      interactions       of      cosmic      ray      electrons
\cite{st70},\cite{st71},\cite{st77}.  Of these processes, the decay of
neutral pions  produced by  cosmic rays interacting  with interstellar
gas   is   expected   to   dominate   at  energies   above   0.1   GeV
\cite{st77},\cite{hu97}.  The  vast   majority  of  these  \grays  are
produced  by   cosmic  rays  with   energies  below  $\sim   20  \GeV$
\cite{st73}.   In   this  energy   range  the  cosmic   ray  spectrum,
particularly in the local  galactic neighborhood that accounts for the
\gray  production  at  high   galactic  latitudes,  is  well  measured
\cite{sh07}. The  pion production cross  section at these  energies is
also very well known \cite{st73},\cite{sa06}.  At \gray energies up to
$\sim$1  GeV, the  pion production  process is  well described  by the
``isobar-plus-fireball''   model  \cite{st70},\cite{de86},\cite{st88}.
At energies  above $\sim$1 GeV,  it is generally assumed  that scaling
holds and that the \gray spectrum will have the same spectral index as
the  primary  proton  spectrum \cite{st88},\cite{st79}.   More  recent
calculations  \cite{mo97}  are  in  good agreement  with  the  results
presented in  Refs.~ \cite{st88} and \cite{st79},  confirming that the
physics  of pion  production at  ~GeV energies  is well  determined. A
modification  of this  physics  involving scaling  violation has  been
suggested \cite{ka05},  however the resulting  effect is too  small to
explain the GeV anomaly by scaling violation alone.

It  has  been suggested  that  the GeV  anomaly  can  be explained  by
postulating that the average primary  proton spectrum in the galaxy is
harder than that observed  locally. With scaling assumed, the required
proton spectral index is $\Gamma = 2.45$ \cite{mo97},\cite{gr97}. Even
with some scaling violation \cite{ka05}, one requires a spectral index
$\Gamma \simeq  2.5$.  The  problem with this  assumption is  that the
local  proton  spectrum has  a  significantly  steeper measured  index
$\Gamma$ =  $2.76 \pm 0.03$  \cite{sh07}.  At galactic  latitudes above
$\sim 20^{o}$ from the plane, the diffuse \grays are produced by these
locally measured protons (and  a smaller number of cosmic-ray $\alpha$
particles  which have  a  similar spectral  index \cite{sh07}).   Even
should  the proton spectrum  in the  inner galaxy  be harder  than the
local spectrum  at high  galactic latitudes, it  is the  steeper local
spectrum  which produces  the high-latitude  pion-decay $\gamma$-rays.
Therefore,  the predicted  spectrum should  be the  canonical  one. In
other  words,  there  should  be  no  GeV  anomaly  at  high  galactic
latitudes.   This  conclusion would  also  result  if  the anomaly  is
strictly produced by a harder cosmic-ray electron spectrum averaged 
over the whole galactic disk~\cite{po98}.

On the contrary, it is generally accepted that the GeV anomaly is {\it
uniform over  the entire  sky}, a result  which was obtained  from the
EGRET data and which was implicit in Figs. 3 and 5 of Ref.~\cite{sr98}
by the  EGRET group.  We will  show here, in a  very quantitative way,
that the anomaly  is seen at {\it all}  galactic latitudes independent
of the line-of-sight density of galactic hydrogen gas. We will further
argue that  such complete isotropy is  what would be  expected if this
anomaly is most likely traceable to the detector itself.

\section{The Celestial Distribution of the GeV Anomaly}

\begin{figure}[h]

\centerline{\psfig{figure=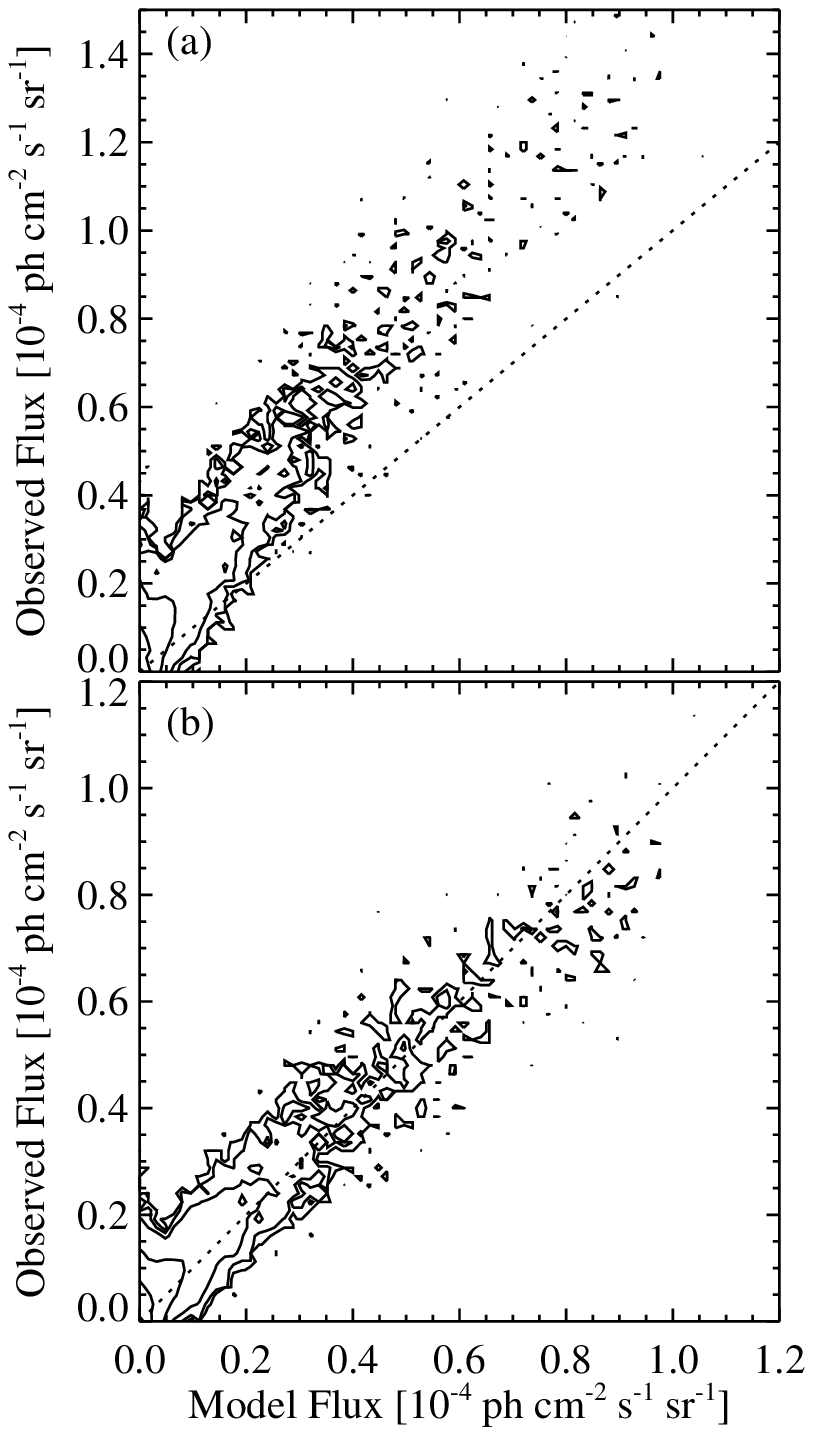,height=6.5in}}

\vspace{2.0cm}
\caption{(a)  Plot of integral ($E  > 1$ GeV),  all-sky diffuse model
flux vs.  EGRET observed flux for  335$^o$ $<$ $l$ $<$ 45$^o$ , $ |b|$
$<$  90$^o$.  (b)  A similar  plot  with a  renormalization factor  of
$(1.6)^{-1}$  applied  to  the  observed  flux.  In  both  plots,  the
integral extragalactic diffuse flux of 1.5 $ \times 10^{-6}$ cm$^{-2}$
s$^{-1}$ sr$^{-1}$  has been added  to the diffuse model.   The dotted
line  indicates the expected  1:1 relationship  between the  model and
observed  fluxes.   Contours  show  the  number  of  0.5$^o$  $\times$
~0.5$^o$ pixels  containing the flux  indicated within a bin  of width
$1.6  \times  10^{-6}$ cm$^{-2}$  s$^{-1}$  sr$^{-1}$  .  The  contour
values  are  $10^4,  10^3,  10^2,  10^1,...   $  These  plots  clearly
demonstrate that the  GeV anomaly exists uniformly over  the whole sky
and extends from high to low intensity galactic flux emission. } 
\label{allsky}
\end{figure}

\begin{figure}[h]

\centerline{\psfig{figure=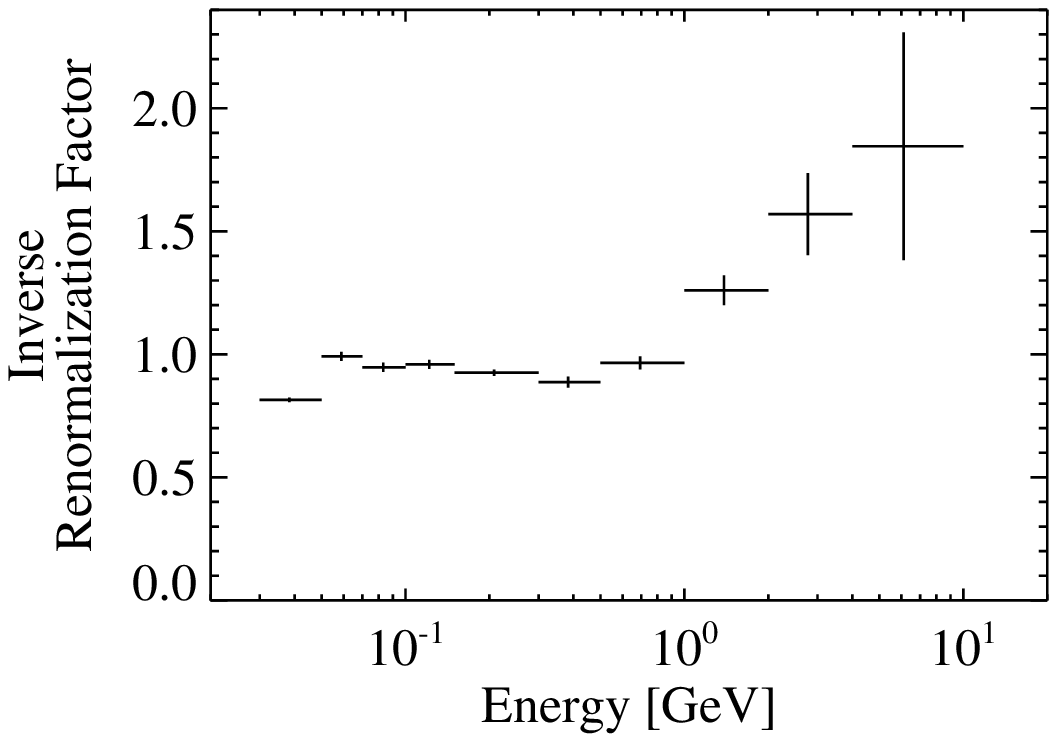,height=4.0in}}

\vspace{2.0cm}

\caption{ Required inverse renormalization  factor for different energy
bins,  given as  the  ratio of  observed-to-predicted  flux {\it  vs.}
energy.}

\label{renorm}
\end{figure}

Our detailed analysis of the galactic \gray flux has now been expanded
to  cover  the  entire  sky~\cite{hu07}.   This  was  accomplished  by
extending the latitude  range of the galactic plane  analysis to $\pm$
25$^o$,  chosen  to  match  the  latitude  range  which  includes  all
significant  emission by  molecular  hydrogen clouds~\cite{da87},  and
further extending the model to the galactic poles~\cite{hu07} by using
the  all-sky Leiden-Dwingaloo \cite{ha97}  and Instituto  Argentino de
Radioastronomía (IAR) Southern Hemisphere HI survey \cite {mo00}.  The
three dimensional cosmic-ray density for the latitude range $ |b|$ $<$
25$^o$ was derived from the  Galactic plane matter distribution on the
assumption  of  dynamic  balance  \cite{be96}.  The  \gray  production
function per H atom in the  22$^o$ $<$ $|b|$ $<$ 25$^o$ latitude range
was extended to  the poles on the assumption  that \gray production in
this range  is dominated by  interactions with cosmic rays  having the
locally  measured  spectrum.   The  analysis  approach  used  in  Ref.
\cite{sr98} to  determine the extragalactic diffuse  emission was then
repeated  using the new  all-sky model  for each  of the  ten standard
energy ranges  as well  as the four  broad energy ranges  discussed in
Ref.  \cite{hu97}.  This analysis, using the EGRET phase 1 and phase 2
source-subtracted data,  yields an extragalactic  \gray spectrum which
is  consistent  with  the  one  derived in  Ref.   \cite{sr98}.   This
consistency is expected because the extragalactic flux is the residual
observed flux as the galactic  model flux is extrapolated to zero. The
slope of the extrapolation does  not affect the intercept, as shown in
Figure~\ref{allsky}. Furthermore, this  all-sky analysis confirms that
the GeV anomaly  is {\it uniform over the entire  sky}, a result which
was implicit in Figs. 3 and  5 Ref.~\cite{sr98}, but which is now more
quantitatively shown  in Fig.~\ref{allsky}. The  lack of
any structure in  the anomaly related to the  Galactic plane, galactic
center, anti-center  or halo, strongly indicates that  the GeV anomaly
is  due to a  systematic error  in the  EGRET calibration  (see below)
rather  than being  a  real astronomical  effect.   The excellent  1:1
correlation of the EGRET data with the expanded all-sky model over the
entire sky and the entire range of emission, after multiplication by a
single renormalization factor, further corroborates this conclusion as
also shown in  Fig.~\ref{allsky}.  Table  1 and Fig.~\ref{renorm} breaks
down  this  effect   as  a  function  of  energy   which  indicates  a
miscalibration which gets worse at the higher energies.

The  overall effect  of applying  this renormalization  correction for
point sources  is slight over a  large energy range on  a log-log plot
and does  not conflict  with any theoretical  models within  the large
observational error bars  of the data above 1  GeV. For \gray ~blazars
observed  by EGRET,  the  fit  to power-law  spectra  would result  in
slightly  steeper spectra,  with  the error  bars  on specific  points
generally large in the energy range above 1 GeV.

\begin{figure}[h]

\centerline{\psfig{figure=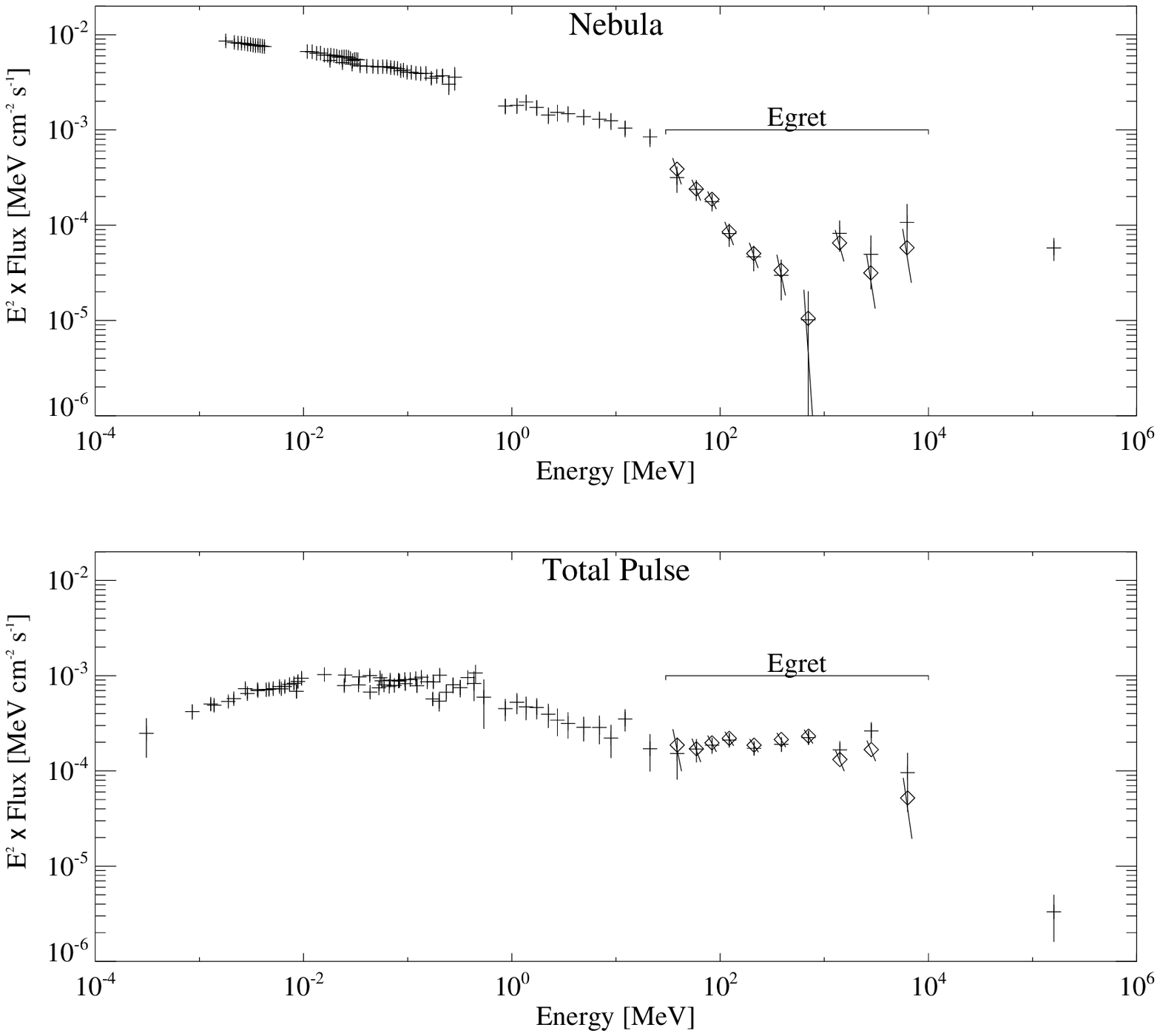,height=6.0in}}

\vspace{2.0cm}

\caption{ Pulsed  and  Nebula spectrum  of  the  Crab  Nebula with  and
without the correction factors from Figure \ref{renorm} applied to the
EGRET data, based on the compilation in Ref. ~\protect\cite{ku01}. The
crosses in  the energy range from 50  MeV to 10 GeV  show the original
EGRET data points and the diamonds show the renormalized values.}
\label{Crab}
\end{figure}

To specifically  illustrate the effect of  our renormalization factors
on the spectra  of point sources, we have  applied our renormalization
to the EGRET  data on both the pulsed and  unpulsed nebula spectrum of
the Crab Nebula as shown in Figure \ref{Crab} based on the compilation
in Ref.~ \cite{ku01}. The renormalized spectrum is consistent with
previous theoretical models of \gray ~emission from the Crab Nebula 
~\cite{ku01},\cite{de96}.

\section{The Dark Matter Hypothesis}

It has  been suggested  that the GeV  anomaly is  the result of  a new
component to  the diffuse \gray spectrum,  one that is  in addition to
the pion-decay component, namely  a component from supersymmetric dark
matter annihilation  \cite{de05}. However, there  are serious problems
with this hypothesis: (1) The same process of dark matter annihilation
would produce  a flux of cosmic-ray antiprotons  which is incompatible
with  the  measured  value   \cite{be06}.   (2)  The  celestial  \gray
distribution  from  dark  matter  annihilation would  be  both  highly
asymmetric  and  clumped \cite{di07}.   This  is again in  direct
contradiction with isotropy of the anomaly.

\section{The Energy Calibration of the EGRET Detector}
 
It was pointed out in  the EGRET calibration paper~\cite{th93} that at
high energies the EGRET sensitivity is poorly known, partly due to the
uncertain effect of self-vetoing caused by charged-particle backsplash
impinging  onto  the  anticoincidence  scintillator  following  shower
generation in  the detector~\cite{sr98}.  This  effect eliminates good
\gray events.  Any uncertainty in the self-veto correction would alter
the sensitivity of the detector at energies above $\sim$ 1 GeV.

An  attempt  to  quantify  the  amount  of  backsplash  was  performed
experimentally up  to 10  GeV during the  calibration of EGRET  at the
Stanford Linear  Accelerator Center (SLAC) with over  a thousand hours
of \gray beam  exposure.  The beam was scanned across  the face of the
tracking detector at a series  of off-axis angles from 0$^o$ to 40$^o$
with three  different roll-axis angles and 10  different energies.  An
EGS (Electron  Gamma Shower) Monte Carlo  (MC) code was  used to study
the  effect of  the  backsplash under  differing  conditions, but  the
fidelity of the MC model was low.  It did indicate that the backsplash
was strongly  dependent on  the charged-particle threshold,  which was
known to be strongly variable across the anti-coincidence detector. The
photomultiplier  gains  were reduced  after  the  SLAC calibration  to
mitigate the higher than  anticipated backsplash effect. The effect of
lowering the  gains is  substantial, and is  not reflected in  the 
sensitivity calibration results as given in Ref.~\cite{th93}.

The  all-sky pervasiveness  of the  apparent excess  flux above  1 GeV
energy (see  above) strongly indicates that this  anomaly is intrinsic
to  the detector.   A systematic  error  in the  understanding of  the
detector  response  would be  one  likely  source.   In an  effort  to
investigate this  possibility, we  have reexamined the  EGRET archival
data carefully,  selecting only the  highest quality EGRET  data.  The
selection  was based  on:  (1)  choosing \gray  events  with the  most
reliable  energy  determination,  (2)  selecting \gray  events  within
30$^o$ of the detector axis  for which the instrument response is most
reliably determined, and (3) excluding \gray events within a 4$\sigma$
angular uncertainty  from the Earth's horizon.  We  have examined data
on the pulsed  and unpulsed emission from the  Crab, Vela, and Geminga
and the diffuse  flux from the Lockman  Hole for phases 1 --  4 of the
EGRET  data  for  energies  $>1  GeV$,  We  find  that  for  different
observations of these sources, there were variations in their measured
fluxes  in excess  of  40\%.  Since  the  flux from  these sources  is
expected  to  be constant,  this  indicates  systematic errors  caused
apparent time  variations over  different observing periods.   This is
not  surprising, since  the effect  of degraded  performance  with the
aging of the spark chamber gas  was calibrated in flight by fixing the
$>$ 100  MeV flux and not  taking account of the  energy dependence of
this effect~\cite{es99}, thus producing a false variability at GeV 
energies.

While we find anomalous behavior pointing to an imprecise knowledge of
the detector  response, we  are unable to  explain the GeV  anomaly as
caused  by  any  single  systematic  instrument effect  that  we  have
studied.   Other contributors  besides an  imprecise knowledge  of the
detector  response that  could explain  an apparent  trans-GeV anomaly
include  (1) an  error in  the  analysis which  might cause  non-\gray
events to be  identified as good high-energy events,  (2) an incorrect
assignment of energies  and (3) an induced detector  background in the
harsh environment of a low-Earth  orbit.  It is difficult to eliminate
any of  these possibilities, but  no specific evidence has  been found
that they  exist. 

Given the systematic  energy dependent uncertainties
in the  sensitivity calibration for the EGRET  instrument discussed in
this  paper,  the published  EGRET  results  above  1 GeV  cannot  be
reliably  depended upon. We  suggest that  one should  incorporate the
energy  dependent  correction factor  shown  in  Figure  2. This 
suggestion applies to possible  use by the GLAST LAT (Large Area
Telescope) collaboration
for comparison  as a sensitivity calibration check  on their extensive
Monte Carlo  simulations and beam tests. We must await 
results from GLAST in order to pursue this question further.
 
\begin{table}
~
\caption{INVERSE NORMALIZATION FACTOR VS.ENERGY} 
\begin{tabular}{|c|l|c|} 
\hline
~
Energy Bin & Inverse Normalization Factor \\
\hline

30-50 MeV  &   0.815 $\pm$ 0.009 \\
50-70 MeV  & 0.992  $\pm$ 0.018 \\ 
70-100 MeV  & 0.947 $\pm$ 0.019 \\  
100-150 MeV &  0.959 $ \pm$ 0.019 \\
150-300 MeV  &  0.925 $\pm$ 0.013 \\  
300-500 MeV & 0.887  $ \pm$ 0.022 \\
500-1000 MeV & 0.965 $\pm$ 0.026 \\
1-2 GeV  & 1.26 $\pm$ 0.06  \\  
2-4 GeV & 1.57  $\pm$ 0.17 \\
4-10 GeV &  1.85 $\pm$ 0.46 \\  

\hline
\end{tabular}

\end{table}

~

\section{Conclusions}

Our new all-sky  analysis of the EGRET data confirms a systematic anomaly
in the form of an  apparent excess galactic emission at energies above
1 GeV over what is expected from galactic cosmic ray interactions with
gas nuclei in atomic and molecular  clouds and interstellar photons,  
as  previously reported~\cite{hu97}. We find further quantitative
support that this anomaly is constant
over  the whole  sky, not  being correlated  with any  astronomical or
galactic features. It is therefore  most likely caused by a systematic
error    in    the    calibration    of   the    effective    detector
sensitivity. Although a detailed reanalysis of the calibration data is
impossible  at this  time,  we have  shown  that plausible  systematic
uncertainties in  the calibration of  the EGRET sensitivity  for \gray
energies above  1 GeV can readily account for the  universal anomalous
excess flux. We  conclude that  neither making
modifications to the observed  primary cosmic-ray nucleon and electron
spectra in order to push  up  the  predicted   theoretical  \gray  flux 
~\cite{st04a}  nor
postulating    a   component    from   supersymmetric    dark   matter
annihilation~\cite{de05}  are  required   to  account  for  the  ``GeV
anomaly''.

Our new analysis is consistent with the extragalactic flux derived in
Ref.~\cite{sr98}. This reevaluation of the EGRET sensitivity also has
implications for calibration checks of the GLAST \gray detector to be 
launched in the near future.

\section*{Acknowledgments}

We thank Michael Salamon and Dave Thompson for their comments and 
suggestions regarding this work
and Zev Gurman for his analysis of EGRET archival data.






\end{document}